\documentclass{article}
\usepackage{spconf,amsmath,graphicx}

\usepackage[colorlinks, linkcolor=black, anchorcolor=black, citecolor=black]{hyperref}
\usepackage{pdfcomment}
\usepackage[pdftex,rgb,dvipsnames,svgnames,hyperref,table]{xcolor}
\usepackage{pgfplots}
\usepackage{caption}
\usepackage{fancyvrb}
\usepackage{amssymb}
\usepackage{amsmath}
\usepackage{amsfonts}
\usepackage{amssymb}
\usepackage{array}
\usepackage{subfigure}
\usepackage{ragged2e}
\usepackage{tabu}
\usepackage{balance}
\usepackage{algorithmicx}
\usepackage{algorithm}
\usepackage{algpseudocode}
\usepackage[marginal]{footmisc}
\usepackage{graphicx}
\usepackage{subfigure}
\usepackage{float}
\usepackage{balance}
\usepackage{cite}
\usepackage{array}
\usepackage{booktabs}
\usepackage{graphicx}
\usepackage{adjustbox}
\usepackage{xcolor}

\title{Cardiac Disease Diagnosis on Imbalanced Electrocardiography Data Through Optimal Transport Augmentation}
\name{\begin{tabular}{c}Jielin Qiu$^{*1}$, Jiacheng Zhu$^{*1}$, Mengdi Xu$^{1}$, Peide Huang$^{1}$, \\ \textit{Michael Rosenberg}$^{3}$, \textit{Douglas Weber}$^{1}$, \textit{Emerson Liu}$^{2}$, \textit{Ding Zhao}$^{1}$\end{tabular}  \thanks{*Equal contribution} }

\address{$^{1}$Carnegie Mellon University, $^{2}$Allegheny Health Network, $^{3}$University of Colorado}

\begin{document}
%
\maketitle

\begin{abstract}
In this paper, we focus on a new method of data augmentation to solve the data imbalance problem within imbalanced ECG datasets to improve the robustness and accuracy of heart disease detection. By using Optimal Transport, we augment the ECG disease data from normal ECG beats to balance the data among different categories. We build a Multi-Feature Transformer (MF-Transformer) as our classification model, where different features are extracted from both time and frequency domains to diagnose various heart conditions. Our results demonstrate 1) the classification models’ ability to make competitive predictions on five ECG categories; 2) improvements in accuracy and robustness reflecting the effectiveness of our data augmentation method. 
\end{abstract}
\section{INTRODUCTION}
The 12-lead Electrocardiogram (ECG) is the foundation for much of cardiology and electrophysiology. It provides unique information about the structure and electrical activity of the heart and also systemic conditions, through changes in timing and morphology of the recorded waveforms. Computer-generated interpretations are standardly provided following ECG acquisition, utilizing predefined rules and algorithmic pattern recognition. However, current approaches miss a lot of the specialized insights and nuances that practiced cardiologists or electrophysiologists can see. Depending on experience, physician reads can also be variable and inconsistent. Achievement of reliable ECG reading would be a significant achievement, where critical and timely ECG interpretations of acute cardiac conditions can lead to efficient and cost-effective intervention.  


With the development of machine learning and deep learning methods, many models have been applied to diagnosing ECG signals \cite{Shanmugam2019MultipleIL}. One of the main issues is that the available data is mostly imbalanced, where the number of labeled ECG signals for a certain condition is very small, so the training samples contain many healthy ECG signals, making it difficult to classify the ECG signals with diseases due to the introduced imbalance in the original data.

In this paper, we propose a new data augmentation method from a probability perspective based on Optimal Transport. We perturb the data distribution towards other classes along the geodesic in a Wasserstein space. Also, the ground metric of this Wasserstein space is computed via a set of physiological features so that the perturbation lies on a manifold that exploits the unique properties of ECG data. We employ a Multi-Feature Transformer as the base classifier to evaluate the performance of our proposed method.

\section{RELATED WORK}
\textbf{Data Imbalance and Data Augmentation in ECG} Traditionally, when facing imbalanced data problems, data augmentation is required before training, aiming to eliminate the effect caused by the imbalance. 
The ECG data imbalance issue has been a long-standing problem. \cite{Martin2021RealtimeFS} tried to use oversampling method to augment the imbalanced data. \cite{ClementVirgeniya2021AND} tried to feed the data into the adaptive synthetic (ADASYN) \cite{He2008ADASYNAS} based sampling model, instead of using synthetic minority oversampling technique (SMOTE) \cite{Chawla2002SMOTESM}. \cite{Liu2021MultiLabelCO} augmented the ECG data by band-pass filter, noise addition, time-frequency transform and data selection. 
The methods above showed that balanced dataset performance is superior to unbalanced one.

\textbf{Optimal Transport (OT)} is a field of mathematics that studies the geometry of probability spaces \cite{Villani2003TopicsIO}. The theoretical importance of OT is that it defines the Wasserstein metric between probability distributions. It reveals a canonical geometric structure with rich properties to be exploited. The earliest contribution to OT originated from Monge in the eighteenth century. Kantorovich rediscovered it under a different formalism, namely the Linear Programming formulation of OT. With the development of scalable solvers, OT is widely applied to many real-world problems \cite{Zhu2021FunctionalOT,Flamary2021POTPO,Zhu2023InterpolationFR}.

\textbf{Machine learning in ECG} With the development of machine learning, many models have been applied to ECG disease detection \cite{Raghunath2021DeepNN,Giudicessi2021ArtificialIA,Strodthoff2021DeepLF}. 
ECG signal can be considered as one type of sequential data, and Seq2seq models \cite{Sutskever2014SequenceTS} are widely used in time series tasks. Since the attention mechanism was proposed \cite{Bahdanau2015NeuralMT}, the Seq2seq model with attention has been improved in various tasks, which outperformed previous methods. The Transformer model \cite{Vaswani2017AttentionIA} achieved even better results, and has recently been adopted in several ECG applications, i.e., arrhythmia classification,  abnormalities detection, stress detection, etc  \cite{Yan2019FusingTM,Che2021ConstrainedTN,Natarajan2020AWA,Behinaein2021ATA,Song2021TransformerbasedSF,Weimann2021TransferLF,zhudata,Zhu2022GeoECGDA,Qiu2023TransferKF}. But those models take only ECG temporal features as input and haven't considered the frequency domain features. 
The OT-based data augmentation is shown to benefit the robustness of ECG diagnosis models~\cite{Zhu2022GeoECGDA}. However, in this work, we focus on the imbalance of datasets and demonstrate the effectiveness of our method, as eliminating imbalance can significantly improve prediction performance.    

\section{METHODS}
\subsection{Overall Pipeline}
\vspace{-10pt}
\begin{figure}[H]
	\centering
	\includegraphics[width=0.49\textwidth]{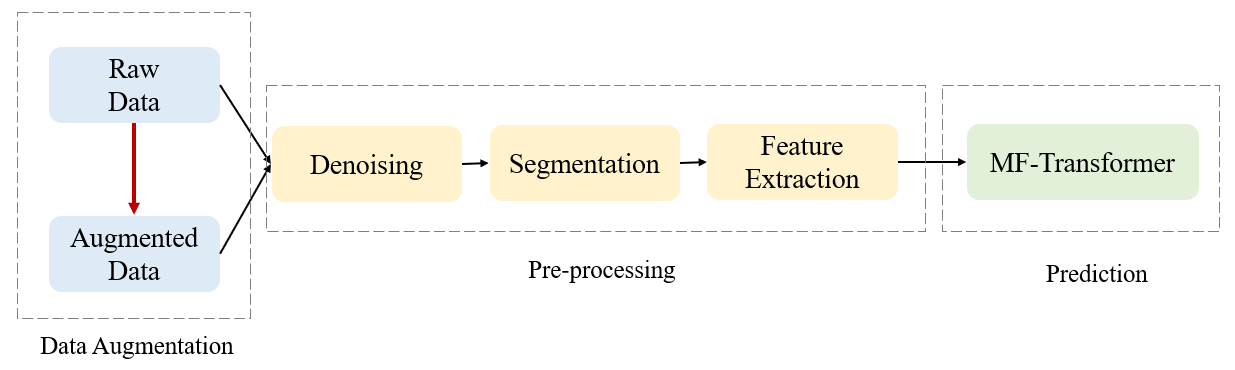}
	\caption{The overall pipeline of the framework.}
	\label{fig:pipeline}
\end{figure}
\vspace{-10pt}
The overall pipeline of our method is shown in Fig. \ref{fig:pipeline}. Given the raw ECG data, we first use Optimal Transport to augment the data of minority categories to solve the data imbalance issue. Then we perform pre-processing on the raw ECG data as well as the augmented data, which includes data denoising, ECG temporal segmentation, and feature extraction. We extract multiple ECG features from both the time domain and frequency domain. Then we use the MF-Transformer as our classification model for performance prediction and evaluation. The details of each part are introduced in Section \ref{MF}, Section \ref{sec_OT}, and Section \ref{pre}, respectively.

\subsection{Multi-Feature Transformer}\label{MF}

\begin{figure}[H]
	\centering
	\includegraphics[width=0.49\textwidth]{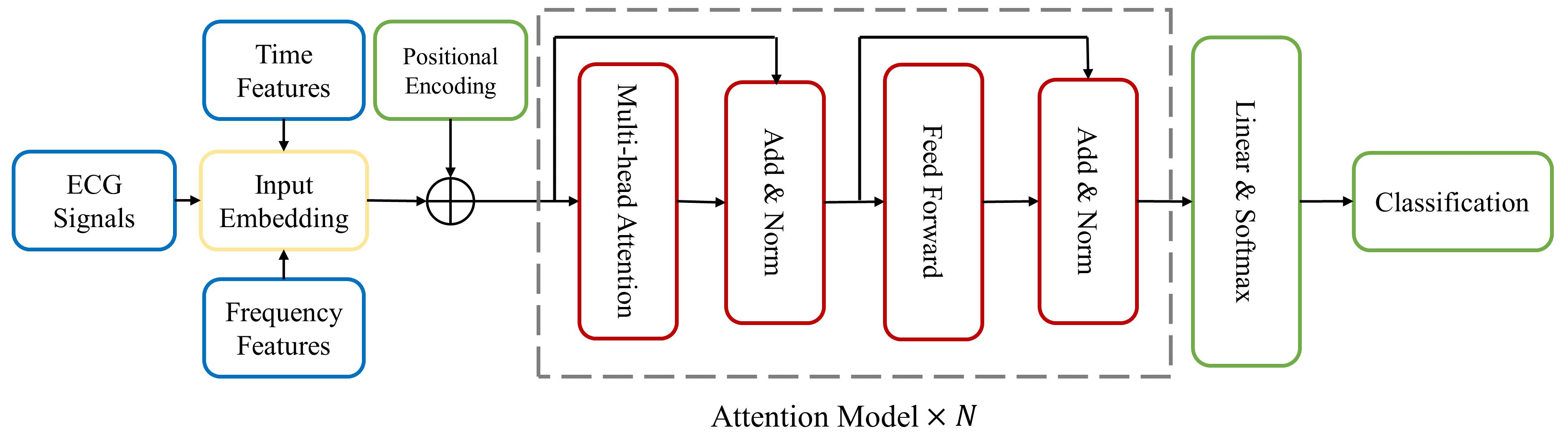}
	\caption{Architecture of the Multi-Feature Transformer model.}
	\label{fig:transformer}
\end{figure}
\vspace{-10pt}
For the classification model, we take advantage of the transformer encoder \cite{Vaswani2017AttentionIA}, and propose a Multi-Feature Transformer (MF-Transformer) model. The transformer is based on the attention mechanism \cite{Vaswani2017AttentionIA} and outperforms previous models in accuracy and performance. The original transformer model is composed of an encoder and a decoder. The encoder maps an input sequence into a latent representation, and the decoder uses the representation along with other inputs to generate a target sequence. Our model is mostly based on the encoder, since we aim to learn the representations of ECG features, instead of decoding them to another sequence. 

The input for the Multi-Feature Transformer is composed of three parts, including ECG raw features, time-domain features, and frequency-domain features. The detailed feature pre-processing steps are introduced in Section~\ref{pre}. First, we feed out the input into an embedding layer, which is a learned vector representation of each ECG feature, by mapping each ECG feature to a vector with continuous values. Then we inject positional information into the embeddings by \cite{Vaswani2017AttentionIA}.
The attention model contains two sub-modules, a multi-headed attention model and a fully connected network. The multi-headed attention computes the attention weights for the input and produces an output vector with encoded information on how each feature should attend to all other features in the sequence. There are residual connections around each of the two sub-layers followed by a layer normalization, where the residual connection means adding the multi-headed attention output vector to the original positional input embedding, which helps the network train by allowing gradients to flow through the networks directly. Multi-headed attention applies a self-attention mechanism, where the input goes into three distinct fully connected layers to create the query, key, and value vectors. The output of the residual connection goes through layer normalization.

\subsection{Optimal Transport Based Data Augmentation}\label{sec_OT}

We use optimal transport to push forward samples from the distribution of a majority class to a minority class. We expect optimal transport to exploit global geometric information so that the synthetic samples match the real samples. 
In specific, we denote the data from a majority class to be $\mathbf{X}_s = \{x_{s,1},...,x_{s,n_s}\} \in \Omega_s$ and the minority class data to be $\mathbf{X}_t = \{x_{t,1},...,x_{t,n_t}\} \in \Omega_t$. We assume that they are subject to distributions $\mathbf{X}_s \sim \mu_s$ and $\mathbf{X}_t \sim \nu_t$, respectively, and we associate empirical measures to data samples:
\begin{equation}
    \hat{\mu_s} = \sum_{i=1}^{n_s} p_{s,i} \delta_{x_{s,i}} {} \text{ ,  } {} \hat{\nu_t} = \sum_{i=1}^{n_t} p_{t,i} \delta_{x_{t,i}},
\end{equation}
where $\delta_x$ is the Dirac function at location $x$ and $p_{i}$ are the probabilities masses associated to the samples.
Solving the optimal transport objective gives us the coupling: 
\begin{equation}
\label{eq:ot_kantorovich}
    \mathbf{\pi}^* = \arg \min_{\mathbf{\pi} \in {\Pi}} \sum_{i=1}^{n_s} \sum_{j=1}^{n_t} \pi_{i,j} C_{i,j} \ + \gamma H(\mathbf{\pi}), 
\end{equation}
where $C_{i,j} = \|x_i - x_j \|^2_2$ is a cost matrix, $\gamma$ is a coefficient, and $H(\pi) = \sum \pi_{i,j} \log \pi_{i,j}$ is the negative entropy regularization that enables us to employ the celebrated Sinkhorn algorithm \cite{cuturi2013sinkhorn}. The solution to problem (\ref{eq:ot_kantorovich}) actually express the barycentric mapping:
\begin{equation}
\label{eq:barycentric_mapping}
    \hat{x}_{s,i} = \arg \min_{x \in \Omega_t} \sum_{j=1}^{n_t} \pi^* (i, j) c(x, x_{t,j}),
\end{equation}
where ${x}_{s,i}$ is source sample and $\hat{x}_{s,i}$ is the resulting mapped sample. When using $l_2$ norm as the cost function, the barycenter has a convenient format that maps the source samples into the convex hull of target samples as  $\mathbf{\hat{X}_s} = n_s \pi^* \mathbf{\hat{X}_t}$. 

\section{DATASET AND REPROCESSING}\label{pre}
We carried out the experiments on the PTB-XL dataset \cite{Wagner2020PTBXLAL}, which contains clinical 12-lead ECG signals of 10-second length. There are five conditions in total, which include Normal ECG (NORM), Myocardial Infarction (MI), ST/T Change (STTC), Conduction Disturbance (CD), and Hypertrophy (HYP). The waveform files are stored in WaveForm DataBase (WFDB) format with 16-bit precision at a resolution of 1$\mu$V/LSB and a sampling frequency of 100Hz. 

\begin{figure}[H]
	\centering
	\includegraphics[width=0.48\textwidth]{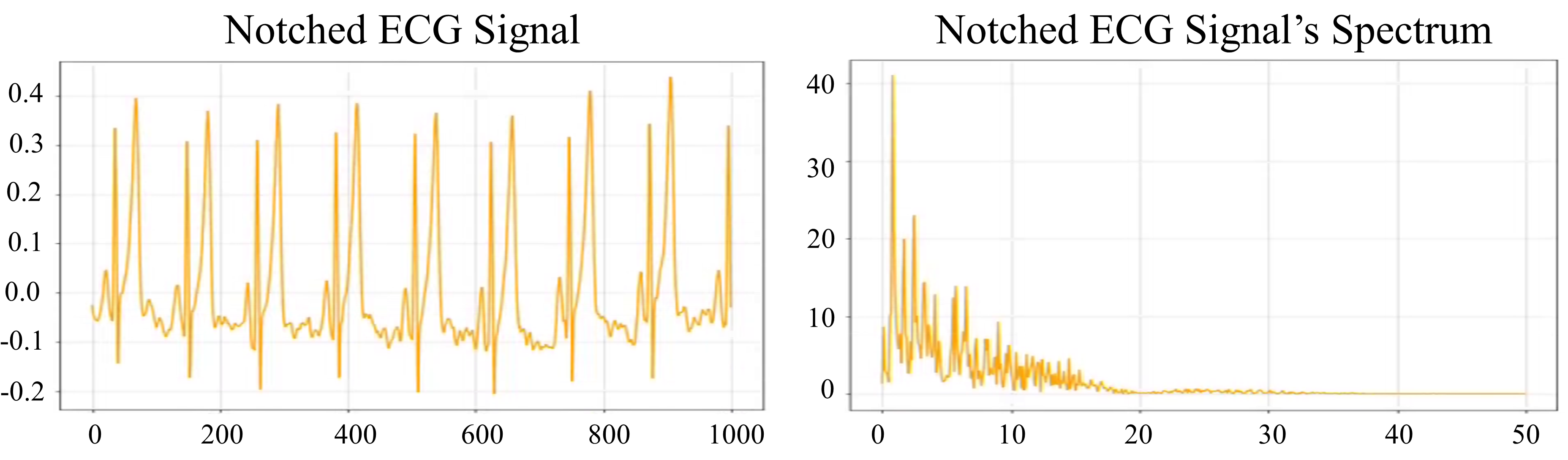}
	\caption{ECG filtered data after n-points window filtering and notch processing.}
	\label{fig:notch}
\end{figure}

\begin{figure}[H]
	\centering
	\includegraphics[width=0.48\textwidth]{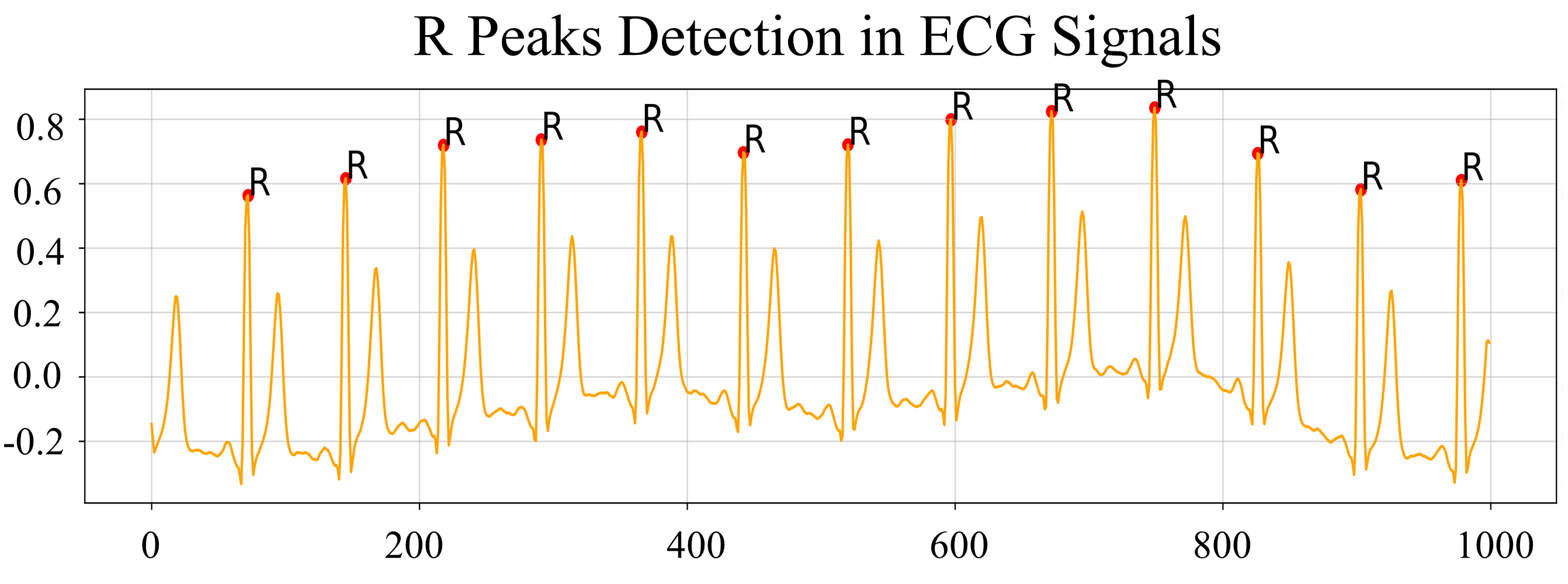}
        \vspace{-5pt}
	\caption{Detecting R peaks in the ECG signals.}
	\label{fig:find_R}
\end{figure}

The raw data was read by wfdb library\footnote{https://pypi.org/project/wfdb/} and Fast Fourier transform (FFT) was performed to process the time series data into the spectrum. Then we perform n-points window filtering to filter the noise and adopt notch processing to filter power frequency interference (noise frequency: 50Hz, quality factor: 30), where the filtered result is shown in Fig.~\ref{fig:notch}.
We then detect the R peaks of each signal by ECG detectors\footnote{https://pypi.org/project/py-ecg-detectors/}, so the data can be sliced at the fixed-sized interval on both sides to obtain individual beats. The examples of detecting R peaks in ECG signals are shown in Fig.~\ref{fig:find_R}. 
\vspace{-20pt}
\vspace{-10pt}
To reduce the dimension of ECG features, we downsample the processed ECG signals to 50Hz. Then we extract more time domain features and frequency domain features to better represent the ECG signals. The time-domain features include: maximum, minimum, range, mean, median, mode, standard deviation, root mean square, mean square, k-order moment and skewness, kurtosis, kurtosis factor, waveform factor, pulse factor, and margin factor. The frequency-domain features include: fft mean, fft variance, fft entropy, fft energy, fft skew, fft kurt, fft shape mean, fft shape std, fft shape skew, fft kurt, which are shown below, respectively. 
\vspace{-10pt}
\begin{equation}\scriptsize
Z_1=\frac{1}{N} \sum_{k=1}^{N} F(k), Z_2 = \frac{1}{N-1} \sum_{k=1}^{N}\left(F(k)-Z_{1}\right)^{2}
\end{equation}
\vspace{-10pt}
\begin{equation}\scriptsize
Z_3 = -1 \times \sum_{k=1}^{N}\left(\frac{F(k)}{Z_{1} N} \log _{2} \frac{F(k)}{Z_{1} N}\right),
Z_4 = \frac{1}{N} \sum_{k=1}^{N}(F(k))^{2}
\end{equation}
\vspace{-10pt}
\begin{equation}\scriptsize
Z_5 = \frac{1}{N} \sum_{k=1}^{N}\left(\frac{F(k)-Z_{1}}{\sqrt{Z_{2}}}\right)^{3}, Z_6=\frac{1}{N} \sum_{k=1}^{N}\left(\frac{F(k)-Z_{1}}{\sqrt{Z_{2}}}\right)^{4}
\end{equation}
\vspace{-10pt}
\begin{equation}\scriptsize
Z_7 = \frac{\sum_{k=1}^{N}(f(k)-F(k))}{\sum_{k=1}^{N} F(k)},
Z_8 =\sqrt{\frac{\sum_{k=1}^{N}\left[\left(f(k)-Z_{6}\right)^{2} F(k)\right]}{\sum_{k=1}^{N} F(k)}}
\end{equation}
\vspace{-10pt}
\begin{equation}\scriptsize
Z_9=\frac{\sum_{k=1}^{N}\left[(f(k)-F(k))^{3} F(k)\right]}{\sum_{k=1}^{N} F(k)},
Z_{10} =\frac{\sum_{k=1}^{N}\left[(f(k)-F(k))^{4} F(k)\right]}{\sum_{k=1}^{N} F(k)}
\end{equation}
\vspace{-20pt}
\begin{table}[H]\small
    \centering
	\caption{Imbalance within the ECG data.}
	\vspace{-10pt}
	\begin{adjustbox}{width=0.85\linewidth}
		\begin{tabular}{c|c|c|c|c}  
			\hline
			Category & Patients & Percentage & ECG beats &Percentage \\ \hline  
			NORM  &9528  &34.2\%  & 28419 &36.6\% \\ \hline
			MI  &5486  &19.7\%  &10959 &14.1\% \\ \hline
			STTC  &5250  &18.9\%   & 8906 &11.5\%   \\ \hline
			CD  &4907  &17.6\%  & 20955  &27.0\%  \\ \hline
			HYP  &2655  &9.5\%  & 8342  &10.8\%  \\ \hline
	\end{tabular}
	\end{adjustbox}
	\label{imbalance_table}
\end{table}
\vspace{-10pt}
Like many other datasets, the imbalance issue is a problem and needs to be solved before further steps. There are five categories in total, including NORM, MI, STTC, CD, and HYP. In a balanced dataset, each category should occupy the same proportion. In the original dataset, the number of patients in the NORM category is much larger than the others. After dividing the ECG signals into individual beats, the portion of each category changed due to heartbeat variance among people. However, if we count the segmented ECG beats and compare different categories' data, the imbalance issue still exists, which is shown in Table \ref{imbalance_table}. As in Table \ref{imbalance_table}, the NORM category and CD category are much larger than the other three categories, making the dataset unbalanced.

\section{Experiments and Discussions}
\subsection{Data Augmentation by Optimal Transport}
As discussed in Section (\ref{sec_OT}), the augmented data generated by Optimal Transport is used to eliminate the data imbalance among different categories.
In specific, 
(1) We always use NORM individual beats as the source and transport the samples from the NORM into each other minor categories. 
(2) In the augmentation procedure, we randomly sample a batch of ECG signals from both the source and target categories and then use formulation in Equation (\ref{eq:barycentric_mapping}) to get the barycentric mapping samples. The labels of augmented samples are set to be the target categories.
(3) We mix the original data and augmented data and then process them for the MF-Transformer following Fig. \ref{fig:pipeline}. We segment the ECG into individual beats, then we obtain the time and frequency statistical features with the method introduced in Section \ref{pre}. After that, we concatenate the ECG signals with all the features as the input for the MF-Transformer model.  
\vspace{-10pt}
\begin{figure}[H]
	\centering
	\includegraphics[width=0.22\textwidth]{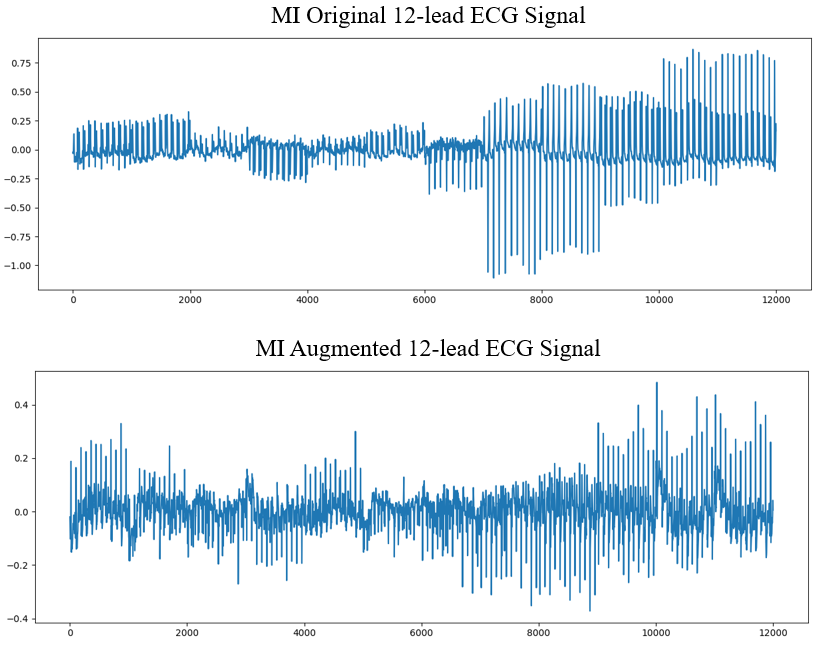}
	\includegraphics[width=0.22\textwidth]{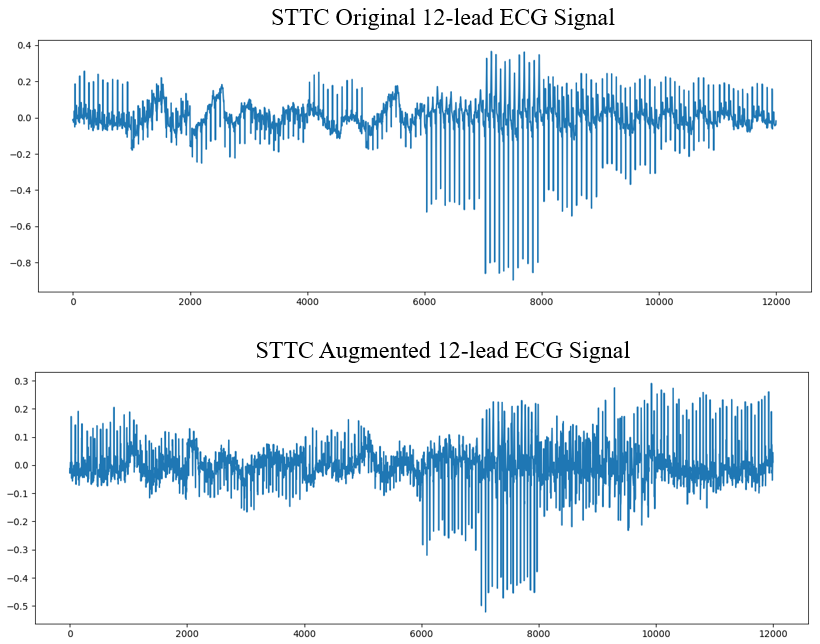}
	\includegraphics[width=0.22\textwidth]{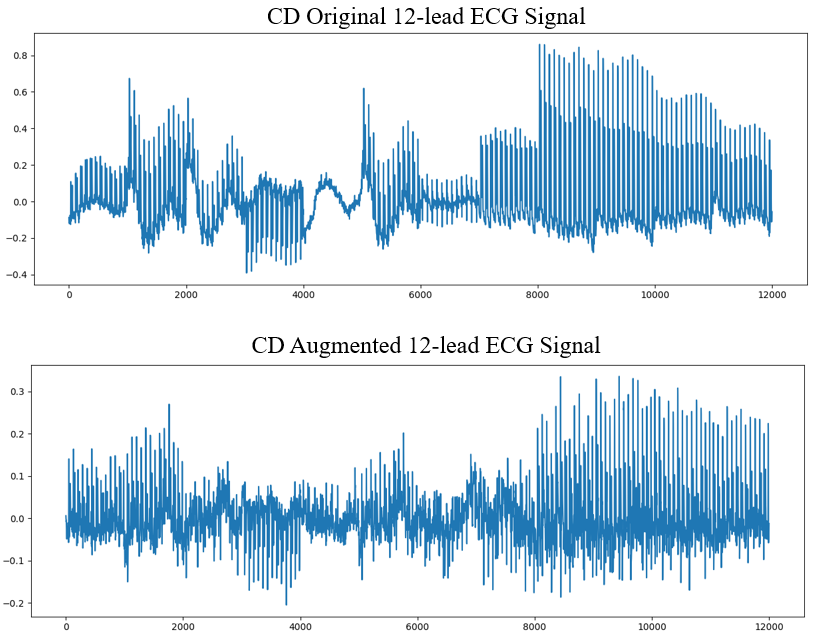}
	\includegraphics[width=0.22\textwidth]{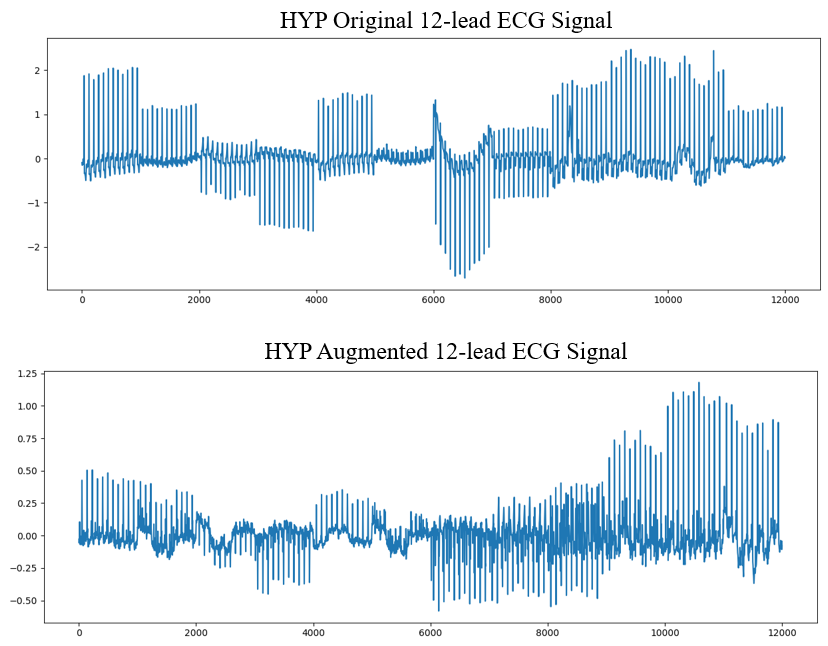}
	\caption{Examples of the original ECG signals and the augmented ECG signals within different conditions. The top row shows the 10-second 12-lead ECG signals of different heart condition categories. The bottom row shows the corresponding transported samples obtained from our Optimal Transport based data augmentation method.}
	\label{fig:aug_result}
\end{figure}
\vspace{-20pt}
\begin{figure}[H]
	\centering
	\includegraphics[width=0.49\textwidth]{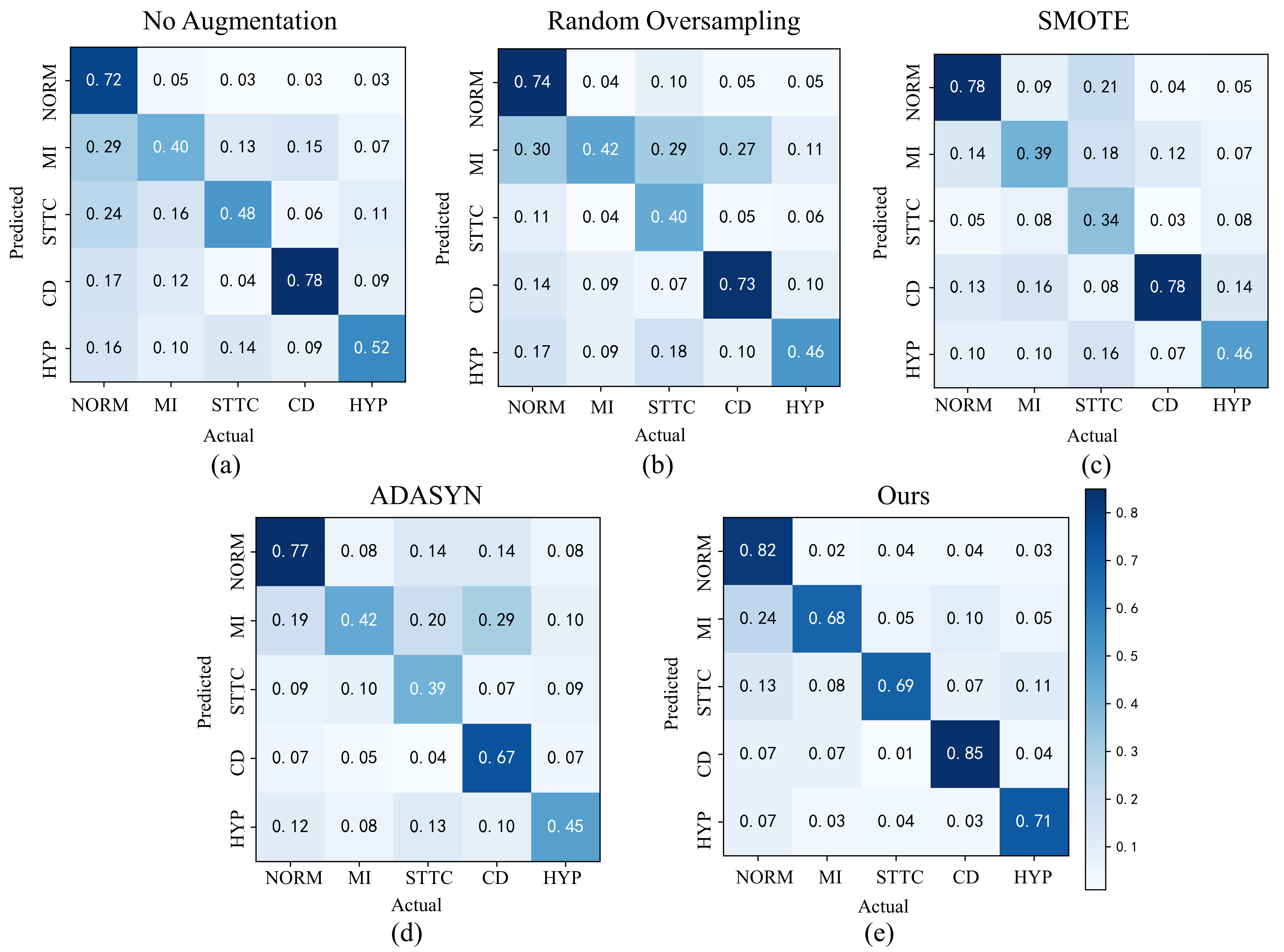}
	\caption{Confusion matrix of prediction results.}
	\label{fig:cm_all}
\end{figure}
\vspace{-20pt}
\subsection{Heart Disease Detection}
To evaluate our methods, we performed experiments on the PTB-XL dataset to predict the category of ECG signals. First, we trained the MF-Transformer model with the original PTB-XL data to obtain the baseline performance for different categories. Second, we used the oversampling strategy to augment the ECG signals for the minority categories, then we trained the MF-Transformer model from scratch to obtain the performance by oversampling data augmentation method. Third, we augmented the data with our OT-based data augmentation method, and trained the MF-Transformer model from scratch again to evaluate the performance of our method. Note that the augmented data is only used for training, and the testing set remains the same as for all the experiments, which only contain the real-world ECG signals to have a fair evaluation of the proposed method. 
The training and testing splitting strategy are the same as in \cite{Wagner2020PTBXLAL,Strodthoff2021DeepLF}. 
The experiments are carried out on four Nvidia Tesla V100 GPUs. The results of different approaches are shown in Fig. \ref{fig:cm_all}, which shows the corresponding accuracy of each category in the confusion matrix. Also, quantitative results in Table \ref{results_table} indicate that our method outperforms all existing baselines. 

\begin{table}[tp]\small
    \centering
	\caption{Comparison of AUROC and F1-score results of heart disease diagnosis by different data augmentation methods.}
	\vspace{-5pt}
        \begin{adjustbox}{width=0.99\linewidth}
		\begin{tabular}{l|cccccccc}  
			\toprule
			Methods & No Aug & Random Oversampling & SMOTE \cite{Chawla2002SMOTESM} & ADASYN  \cite{He2008ADASYNAS} & TaskAug \cite{raghu22a} & GeoECG \cite{Zhu2022GeoECGDA} & Ours \\
                 \midrule
                AUROC  &0.843  &0.820  &0.799  &0.820 &0.842 &0.931 &\textbf{0.944} \\
                F1-score  &0.575  &0.536  & 0.534 &0.546 &-- & 0.707 &\textbf{0.738} \\
                \bottomrule
	\end{tabular}
       \end{adjustbox}
	\label{results_table}
 \vspace{-10pt}
\end{table}

From Fig.~\ref{fig:cm_all}(a), we can see that due to the imbalance within the dataset, the accuracies of NORM and CD categories are much higher than the other three, meaning the model learns more patterns from those two categories compared with the others, which could have a negative impact since the other three heart disease categories can only achieve around 50\% $\sim$ 60\%  classification accuracy. So the data imbalance issue needs to be resolved to improve the model's performance and robustness to different heart conditions.

Fig.~\ref{fig:cm_all}(b) shows by oversampling data from minority categories to make the data more balanced, the classification accuracy increased, especially for the minority categories. But the improvement is not high enough, and could easily lead to overfitting.

Fig.~\ref{fig:cm_all}(c) shows the results by learning from both raw data and our OT-augmented data.  Compared with the oversampling results shown in Fig. \ref{fig:cm_all}(b), we can see that not only the classification accuracy of each category has improved, but the average classification result has also increased from 71.80\% (original) and 72.05\% (oversampling) to 75.82\% (ours). Each category's performance comes to be more balanced, showing the robustness improvement compared with the baseline results and oversampling results in Fig.~\ref{fig:cm_all}(a) and Fig.~\ref{fig:cm_all} (b).

\section{Conclusions and Future Work}
In this paper, we proposed a new method to deal with the ECG data imbalance problem. We augmented the minority category from the majority category with Optimal Transport to make the data balanced, which can handle the overfitting issue introduced by the traditional sampling method. We also proposed an MF-Transformer as our classification model to predict heart conditions from ECG signals. We showed that after data augmentation, there are both accuracy and robustness improvements in the classification results over five ECG categories, which demonstrates the effectiveness of our method.

\clearpage
\section{Acknowledgements}

The research is partially supported by Allegheny Health Network and Mario Lemieux Center for Innovation and Research in EP, and DARPA ADAPTER program.

\small{
\bibliographystyle{IEEEbib}
\bibliography{reference}

\begin{thebibliography}{10}

\bibitem{Shanmugam2019MultipleIL}
Divya Shanmugam, Davis Blalock, and John Guttag,
\newblock ``Multiple instance learning for ecg risk stratification,''
\newblock in {\em Proceedings of the 4th Machine Learning for Healthcare
  Conference}. 2019, vol. 106 of {\em Proceedings of Machine Learning
  Research}, pp. 124--139, PMLR.

\bibitem{Martin2021RealtimeFS}
Harold Martin, Ulyana Morar, Walter Izquierdo, Mercedes Cabrerizo, Anastasio
  Cabrera, and Malek Adjouadi,
\newblock ``Real-time frequency-independent single-lead and single-beat
  myocardial infarction detection.,''
\newblock {\em Artificial intelligence in medicine}, vol. 121, pp. 102179,
  2021.

\bibitem{ClementVirgeniya2021AND}
S.~ClementVirgeniya and E.~Ramaraj,
\newblock ``A novel deep learning based gated recurrent unit with extreme
  learning machine for electrocardiogram (ecg) signal recognition,''
\newblock {\em Biomed. Signal Process. Control.}, vol. 68, pp. 102779, 2021.

\bibitem{He2008ADASYNAS}
Haibo He, Yang Bai, Edwardo~A. Garcia, and Shutao Li,
\newblock ``Adasyn: Adaptive synthetic sampling approach for imbalanced
  learning,''
\newblock {\em IEEE International Joint Conference on Neural Networks}, pp.
  1322--1328, 2008.

\bibitem{Chawla2002SMOTESM}
N.~Chawla, K.~Bowyer, Lawrence~O. Hall, and W.~Philip Kegelmeyer,
\newblock ``Smote: Synthetic minority over-sampling technique,''
\newblock {\em J. Artif. Intell. Res.}, vol. 16, pp. 321--357, 2002.

\bibitem{Liu2021MultiLabelCO}
Yamin Liu, Hanshuang Xie, Qineng Cao, Jiayi Yan, Fan Wu, Huaiyu Zhu, and Yun
  Pan,
\newblock ``Multi-label classification of multi-lead ecg based on deep 1d
  convolutional neural networks with residual and attention mechanism,''
\newblock {\em 2021 Computing in Cardiology (CinC)}, vol. 48, pp. 1--4, 2021.

\bibitem{Villani2003TopicsIO}
C{\'e}dric Villani,
\newblock ``Topics in optimal transportation,''
\newblock 2003.

\bibitem{Zhu2021FunctionalOT}
Jiacheng Zhu, Aritra Guha, Mengdi Xu, Yingchen Ma, Rayleigh Lei, Vincenzo
  Loffredo, XuanLong Nguyen, and Ding Zhao,
\newblock ``Functional optimal transport: Mapping estimation and domain
  adaptation for functional data,''
\newblock {\em ArXiv}, vol. abs/2102.03895, 2021.

\bibitem{Flamary2021POTPO}
R{\'e}mi Flamary et~al.,
\newblock ``Pot: Python optimal transport,''
\newblock 2021.

\bibitem{Zhu2023InterpolationFR}
Jiacheng Zhu, Jielin Qiu, Aritra Guha, Zhuolin Yang, XuanLong Nguyen, Bo~Li,
  and Ding Zhao,
\newblock ``Interpolation for robust learning: Data augmentation on
  geodesics,''
\newblock {\em ArXiv}, vol. abs/2302.02092, 2023.

\bibitem{Raghunath2021DeepNN}
Sushravya Raghunath et~al.,
\newblock ``Deep neural networks can predict new-onset atrial fibrillation from
  the 12-lead ecg and help identify those at risk of atrial
  fibrillation–related stroke,''
\newblock {\em Circulation}, vol. 143, pp. 1287 -- 1298, 2021.

\bibitem{Giudicessi2021ArtificialIA}
John~R. Giudicessi et~al.,
\newblock ``Artificial intelligence-enabled assessment of the heart rate
  corrected qt interval using a mobile electrocardiogram device.,''
\newblock {\em Circulation}, 2021.

\bibitem{Strodthoff2021DeepLF}
Nils Strodthoff, Patrick Wagner, Tobias Schaeffter, and Wojciech Samek,
\newblock ``Deep learning for ecg analysis: Benchmarks and insights from
  ptb-xl,''
\newblock {\em IEEE Journal of Biomedical and Health Informatics}, vol. 25, pp.
  1519--1528, 2021.

\bibitem{Sutskever2014SequenceTS}
Ilya Sutskever, Oriol Vinyals, and Quoc~V. Le,
\newblock ``Sequence to sequence learning with neural networks,''
\newblock in {\em NIPS}, 2014.

\bibitem{Bahdanau2015NeuralMT}
Dzmitry Bahdanau, Kyunghyun Cho, and Yoshua Bengio,
\newblock ``Neural machine translation by jointly learning to align and
  translate,''
\newblock {\em CoRR}, vol. abs/1409.0473, 2015.

\bibitem{Vaswani2017AttentionIA}
Ashish Vaswani, Noam~M. Shazeer, Niki Parmar, Jakob Uszkoreit, Llion Jones,
  Aidan~N. Gomez, Lukasz Kaiser, and Illia Polosukhin,
\newblock ``Attention is all you need,''
\newblock {\em ArXiv}, vol. abs/1706.03762, 2017.

\bibitem{Yan2019FusingTM}
Genshen Yan, Shen Liang, Yanchun Zhang, and Fan Liu,
\newblock ``Fusing transformer model with temporal features for ecg heartbeat
  classification,''
\newblock {\em BIBM}, pp. 898--905, 2019.

\bibitem{Che2021ConstrainedTN}
Chao Che, Peiliang Zhang, Min Zhu, Yue Qu, and Bo~Jin,
\newblock ``Constrained transformer network for ecg signal processing and
  arrhythmia classification,''
\newblock {\em BMC Medical Informatics and Decision Making}, vol. 21, 2021.

\bibitem{Natarajan2020AWA}
Annamalai Natarajan, Yale Chang, Sara Mariani, Asif Rahman, Gregory Boverman,
  Shruti~Gopal Vij, and Jonathan Rubin,
\newblock ``A wide and deep transformer neural network for 12-lead ecg
  classification,''
\newblock {\em 2020 Computing in Cardiology}, pp. 1--4, 2020.

\bibitem{Behinaein2021ATA}
Behnam Behinaein, Anubha Bhatti, Dirk Rodenburg, Paul~C. Hungler, and Ali
  Etemad,
\newblock ``A transformer architecture for stress detection from ecg,''
\newblock {\em 2021 International Symposium on Wearable Computers}, 2021.

\bibitem{Song2021TransformerbasedSF}
Yonghao Song, Xueyu Jia, Lie Yang, and Longhan Xie,
\newblock ``Transformer-based spatial-temporal feature learning for eeg
  decoding,''
\newblock {\em ArXiv}, vol. abs/2106.11170, 2021.

\bibitem{Weimann2021TransferLF}
Kuba Weimann and Tim O.~F. Conrad,
\newblock ``Transfer learning for ecg classification,''
\newblock {\em Scientific Reports}, vol. 11, 2021.

\bibitem{zhudata}
Jiacheng Zhu, Jielin Qiu, Zhuolin Yang, Michael Rosenberg, Emerson Liu, Bo~Li,
  and Ding Zhao,
\newblock ``Data augmentation via wasserstein geodesic perturbation for robust
  electrocardiogram prediction,''
\newblock .

\bibitem{Zhu2022GeoECGDA}
Jiacheng Zhu, Jielin Qiu, Zhuolin Yang, Douglas Weber, Michael~A. Rosenberg,
  Emerson Liu, Bo~Li, and Ding Zhao,
\newblock ``Geoecg: Data augmentation via wasserstein geodesic perturbation for
  robust electrocardiogram prediction,''
\newblock {\em ArXiv}, vol. abs/2208.01220, 2022.

\bibitem{Qiu2023TransferKF}
Jielin Qiu, William Han, Jiacheng Zhu, Mengdi Xu, Michael Rosenberg, Emerson
  Liu, Douglas Weber, and Ding Zhao,
\newblock ``Transfer knowledge from natural language to electrocardiography:
  Can we detect cardiovascular disease through language models?,''
\newblock {\em ArXiv}, vol. abs/2301.09017, 2023.

\bibitem{cuturi2013sinkhorn}
Marco Cuturi,
\newblock ``Sinkhorn distances: Lightspeed computation of optimal transport,''
\newblock {\em Advances in neural information processing systems}, vol. 26, pp.
  2292--2300, 2013.

\bibitem{Wagner2020PTBXLAL}
Patrick Wagner, Nils Strodthoff, R.~Bousseljot, D.~Kreiseler, F.~Lunze,
  W.~Samek, and T.~Schaeffter,
\newblock ``Ptb-xl, a large publicly available electrocardiography dataset,''
\newblock {\em Scientific Data}, vol. 7, 2020.

\bibitem{raghu22a}
Aniruddh Raghu, Divya Shanmugam, Eugene Pomerantsev, John Guttag, and Collin~M
  Stultz,
\newblock ``Data augmentation for electrocardiograms,''
\newblock in {\em Proceedings of the Conference on Health, Inference, and
  Learning}. 2022, Proceedings of Machine Learning Research, pp. 282--310,
  PMLR.

\end{thebibliography}
}
\end{document}